# Emission Characteristics of Energetic Electrons with Crescent-shaped Velocity Distributions

Mehdi Yousefzadeh
Institute of Frontier and Interdisciplinary Science, Shandong University, Qingdao, Shandong 266237, People's Republic of China; yousefzadeh@sdu.edu.cn, m.yousefzadeh6@gmail.com

## Abstract

Solar flares release magnetic energy through reconnection, accelerating electrons into nonthermal velocity distributions, including crescent-shaped electron populations. These energetic electron distributions are crucial in driving instabilities that can lead to distinct electromagnetic emissions. This study investigates the emission properties of crescent-shaped electron velocity distribution functions under different frequency ratios ($\omega_{pe}/\Omega_{ce}$), critical for understanding plasma conditions in various astrophysical environments, by comparing the emissions and intensities of waves among different cases. Here, we study and analyze three distinct frequency ratio conditions (2.2, 10, and 1, designated as cases A, B, and C, respectively). We find that the beam-Langmuir and upper-hybrid modes can be efficiently excited, leading to further plasma emissions in different cases. Our study reveals that the fundamental (O/F) emission can reach a maximum value of $\sim 10^{-4} E_{k0}$, while the harmonics (H) can extend to $\sim 1.5 \times 10^{-5} E_{k0}$, depending on the frequency ratio of the environment. The intensity of the fundamental mode exceeds previous findings for pure-ring, pure-beam, and ring–beam distributions, highlighting the impact of crescent-shaped electron velocity distributions on wave excitation and emission processes. This effect is notably influenced by different frequency ratios, offering new insights into the way that nonthermal electron distributions affect the plasma emission process.

*Unified Astronomy Thesaurus concepts:* Solar corona (1483); Solar activity (1475); Radio bursts (1339); Solar coronal radio emission (1993)

*Materials only available in the online version of record:* animations

## 1. Introduction

The particle acceleration during magnetic reconnection results in nonthermal electron velocity distribution functions (EVDFs). These nonthermal EVDFs can generate microscopic plasma instabilities and emit radio emissions. The shape of an EVDF is influenced by the surrounding electromagnetic field structures and disturbances. For example, the ring–beam distribution, typically formed by injecting beams obliquely into the magnetic field, contains components that drive both the bump-on-tail instability and the electron cyclotron maser instability (ECMI; Y. Chen et al. 2022a, 2022b).

Crescent-shaped EVDFs show a unique shape that has been seen by observation and confirmed through simulations (N. Bessho et al. 2016; X. Yao 2022a). Based on linear dispersion analyses, this type of EVDF has been shown to generate high-frequency electrostatic upper-hybrid (UH) waves characterized by wave oscillations around 90°, as well as beam-mode waves, resulting from beam–plasma interactions (J. Burch et al. 2019). The Magnetospheric Multiscale (MMS) has detected these waves during various events and and at various locations, including the Earth's magnetotail reconnection and the asymmetric magnetopause reconnection (D. B. Graham et al. 2018; K. Dokgo et al. 2019). N. Bessho et al. (2016) and G. Lapenta et al. (2017) have studied and addressed the origin of crescent-shaped VDFs, mainly by MMS observations and MHD and particle-in-cell (PIC) simulations. They have suggested that the electrons of meandering orbits in the diffusion region result in the formation of crescent-shaped EVDFs.

The theoretical framework for plasma emission, established by V. L. Ginzburg & V. V. Zhelezniakov (1958), involves a multistage nonlinear process, where the kinetic bump-on-tail instability induced by energetic electron beams excites the beam-Langmuir (BL) mode, leading to fundamental (F) and harmonic (H) emissions through nonlinear wave–wave interactions (J. P. Wild 1950; J. P. Wild & L. L. McCready 1950; J. P. Wild et al. 1954). Y. Chen et al. (2022a) expanded on this by exploring wave excitation and nonlinear interactions from ring-shaped VDFs, identifying that ECMI and anisotropic instability develop in these systems, producing F and H emissions by coupling different wave modes. They proposed that the F emission is produced by the nonlinear coupling of the slow extraordinary (Z) and the low-frequency Whistler (W) mode, while the H mode is excited by the nonlinear coupling of the primary and the scattered UH mode. These results are in agreement with the earlier works using Dory–Guest–Harris (R. A. Dory et al. 1965) VDFs (S. Ni et al. 2020). Crescent-shaped EVDFs consist of two components, like ring–beam distribution: the beam and the crescent. Each component can initiate a distinct type of instability within the system.

Intrinsic turbulent Alfvén waves have been shown to significantly alter electron motion along the magnetic field through pitch-angle scattering, a process that might lead to the formation of crescent-shaped VDFs in the solar corona (Q. M. Lu et al. 2006). Later studies used test particle simulations to explore how these Alfvén waves impact an electron beam, showing that they contribute to the development of crescent-shaped VDFs. They indicate that Alfvén waves particularly enhance the excitation of the fundamental O mode, especially when the frequency ratio is around unity (C. S. Wu et al. 2012).

X. Yao (2022a) studied the EVDFs generated during 3D magnetic reconnection using fully kinetic PIC simulations. Their research indicates the presence of crescent-shaped EVDFs in certain areas during the reconnection process. These locations were further studied to understand the wave emission process resulting from these particular EVDFs. X. Yao (2022b) reported that crescent-shaped EVDFs can generate multiple-harmonic electromagnetic waves through ECMI in the diffusion region of reconnection.

In X. Yao (2022b), the mass ratio in the simulations was set to 100, which is much below the realistic mass ratio. This was probably done because they investigated the emission properties of the crescent-shaped EVDF following their earlier research on 3D magnetic reconnection, where a mass ratio of 100 was also used (X. Yao 2022a). They also chose a small number of macroparticles per cell per species (∼100). Numerous authors—including Z. Zhang et al. (2022), J. O. Thurgood & D. Tsiklauri (2015), and P. Henri et al. (2019)—have conducted critical reviews emphasizing the importance of ensuring a sufficient number of macroparticles per cell per species to achieve accurate results in PIC simulations. While many studies have employed values below a few hundred due to economic and computational constraints (e.g., T. Umeda 2010), this can introduce significant numerical noise, adversely affecting the signal-to-noise ratio of the captured F/H emissions. The density ratio of energetic electrons to background electrons in their initial configuration setup was set to 0.5. They considered a comparatively high beam-to-background density ratio that could potentially occur only in a localized area during magnetic reconnection simulations (X. Yao 2022a).

To get a better understanding, we have adjusted the mass ratio to a realistic value, significantly increased the number of macroparticles per cell, and modified the density ratio to further analyze the emission properties of the crescent-shaped EVDFs under a similar frequency ratio. Note that the impacts of different frequency ratios ($\omega_{pe}/\Omega_{ce}$) on the emission characteristics of crescent-shaped EVDFs are not well understood. It is also unclear how these conditions affect the excitation of various modes. To provide insights into these fundamental issues, we aim to compare and analyze the wave emissions and intensities in different cases through PIC simulations.

In the next section, we describe the numerical technique of PIC simulations and the parameter setup used to describe the crescent-shaped EVDFs. Section 3 presents the wave analyses and major results for different cases. In Section 4, we provide a summary and discussion of the outcomes, including some comparisons with earlier studies.

## 2. Numerical Model and Parameter Setup

We utilize the vector PIC (VPIC) open-source software provided by Los Alamos National Labs for our simulations, running on the supercomputers of the Beijing Super-Cloud Computing Center.[1] The VPIC code operates in a 2D3V framework, which uses a second-order explicit leapfrog algorithm to update the positions and velocities of particles. This code employs a second-order finite-difference time-domain solver to evolve the electromagnetic fields according to the full Maxwell equations (K. J. Bowers et al. 2008a, 2008b, 2009).

To start the PIC simulation, we use a Maxwellian distribution to represent the background plasma, which includes electrons and protons. The background magnetic field is along the z-direction ($B_0 \hat{e}_z$), while the wavevector ($\mathbf{k}$) lies on the $xOz$ plane. The ratio between energetic and background electrons ($n_e/n_0$) is assumed to be 0.01. The temperature of the background electrons and protons is considered to be around 2 Mk. The grid spacing ($\Delta$) is approximately ∼1.25 times the Debye length of the background electrons ($\lambda_D$), the time step is ∼0.012 $\omega_{pe}^{-1}$, and the grid number is set to be [1024, 1024]. The ratio between the plasma oscillation frequency and the electron gyrofrequency ($\omega_{pe}/\Omega_{ce}$) varies in different cases (i.e., $\omega_{pe}/\Omega_{ce} = 2.2$, 10, and 1 in cases A, B, and C, respectively). These ratios correspond to various conditions observed in phenomena such as the solar corona and the magnetic field of the Earth's magnetopause, where crescent-shaped EVDFs have been observed. The simulation lasts for 2000 $\omega_{pe}^{-1}$ in different cases. In each cell for each type of particle, we use 1000 macroparticles and maintain the charge neutrality in the system. We employ the periodic boundary conditions.

The following equations describe the analytical expression for the crescent-shaped distribution function of energetic electrons in parallel and perpendicular directions (see Figure 1):

$$f_\parallel = A_0 \exp\left[-\left(\frac{v_\parallel - u_{d\parallel}}{\sqrt{2}\, v_{\text{th}}}\right)^2\right],\ f_\perp$$
$$= A_1 v_\perp \exp\left[-\left(\frac{v_\perp - u_{d\perp}}{\sqrt{2}\, v_{\text{th}}}\right)^2 - \left(\frac{\phi - \phi_0}{\sqrt{2}\, \phi_{\text{th}}}\right)^2\right].$$

The speeds of perpendicular and parallel drift are assumed to be $0.2c$, which approximately equals the kinetic energy of around ∼10 keV. Here, $A_0$ and $A_1$ stand for the normalization factor and $\phi$ is the polar angle, which is defined as the arc tangent of the ratio of two perpendicular (orthogonal) components of velocity ($v_{\perp 1}$ and $v_{\perp 2}$). The angular width of the thermal VDF, denoted as $\phi_{\text{th}}$, is specified to be $0.6\pi$, centered around $\phi_0 = 0$, which is similar to the typical crescent-shaped VDF found in the diffusion region (X. Yao 2022a). The product of the above two equations results in a perpendicular crescent-shaped distribution.

## 3. Numerical Results and Analyses

In this study, we explore three cases, labeled A ($\omega_{pe}/\Omega_{ce} = 2.2$), B ($\omega_{pe}/\Omega_{ce} = 10$), and C ($\omega_{pe}/\Omega_{ce} = 1$). All other simulation parameters remain constant across these cases. We start our analysis of the excited wave modes with case A, which reflects the diffusion and outflow region of reconnection observed in the solar corona (X. Yao 2022a; X. Yao 2022b). We investigate wave characteristics and examine the resonance conditions involving wave–particle and wave–wave interactions, then we compare them with the other two cases.

### 3.1. Wave Analysis for Case A

We start from case A, which corresponds to the defined frequency ratio of $\omega_{pe}/\Omega_{ce} = 2.2$. The upper and lower panels of Figure 2(a) show the final stage ($t \sim 2000\, \omega_{pe}^{-1}$) of the PIC-evolved distribution in the $v_\perp - v_\parallel$, and $v_{1\perp} - v_{2\perp}$ planes, respectively. The associated movie illustrates the evolution of this distribution over time. Initially, from 0 to 150 $\omega_{pe}^{-1}$, there is a fast movement of energetic electrons toward lower velocities in the parallel direction, indicating fast deceleration, with no movement perpendicular to the magnetic field. Between 150 and 1100 $\omega_{pe}^{-1}$, the electrons gradually tend to spread out from the parallel to the perpendicular direction toward lower velocities, while gradually there is less movement in the parallel direction. From 1100 to 2000 $\omega_{pe}^{-1}$, the electrons evolve mainly in their initial locations, resulting in almost no further changes in their VDFs. The perpendicular and parallel stages of evolution mainly result from different parts of VDF, influenced respectively by the crescent and beam components.

---

[1] http://www.blsc.cn/

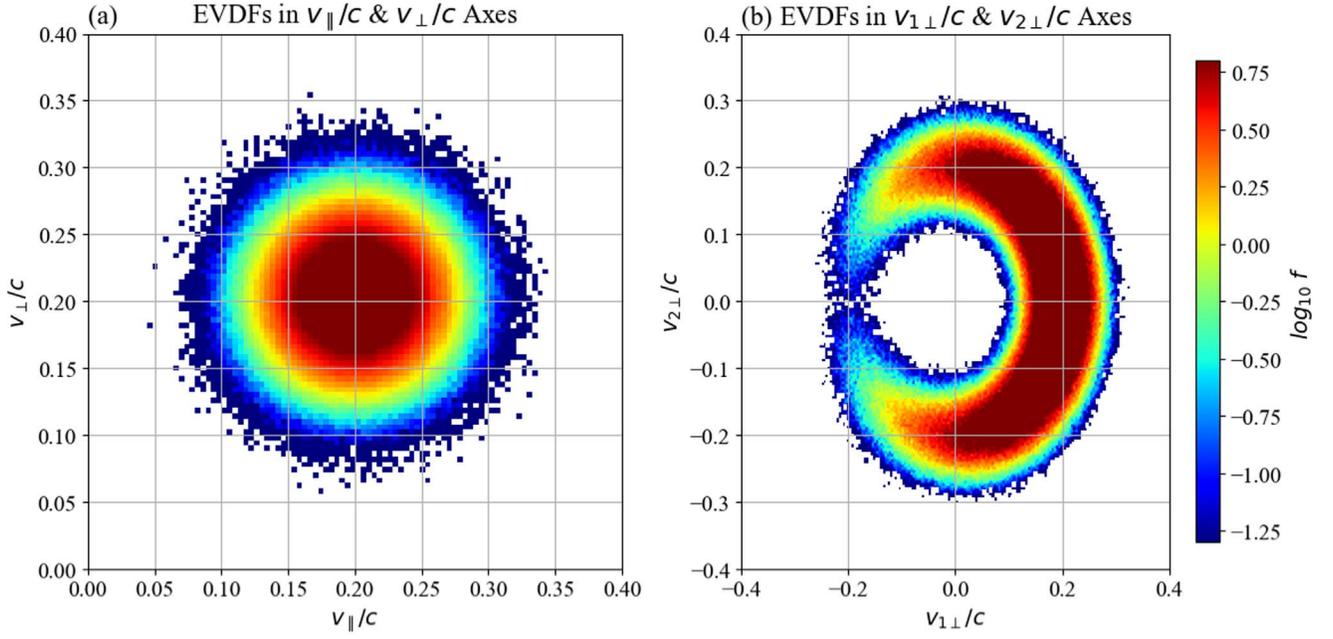

**Figure 1.** Initial crescent-shaped EVDFs of our simulations. The EVDF is depicted in both (a) $v_\perp - v_\parallel$ and (b) $v_{1\perp} - v_{2\perp}$ planes. These shapes are generated by the defined equations for a crescent-shaped distribution.

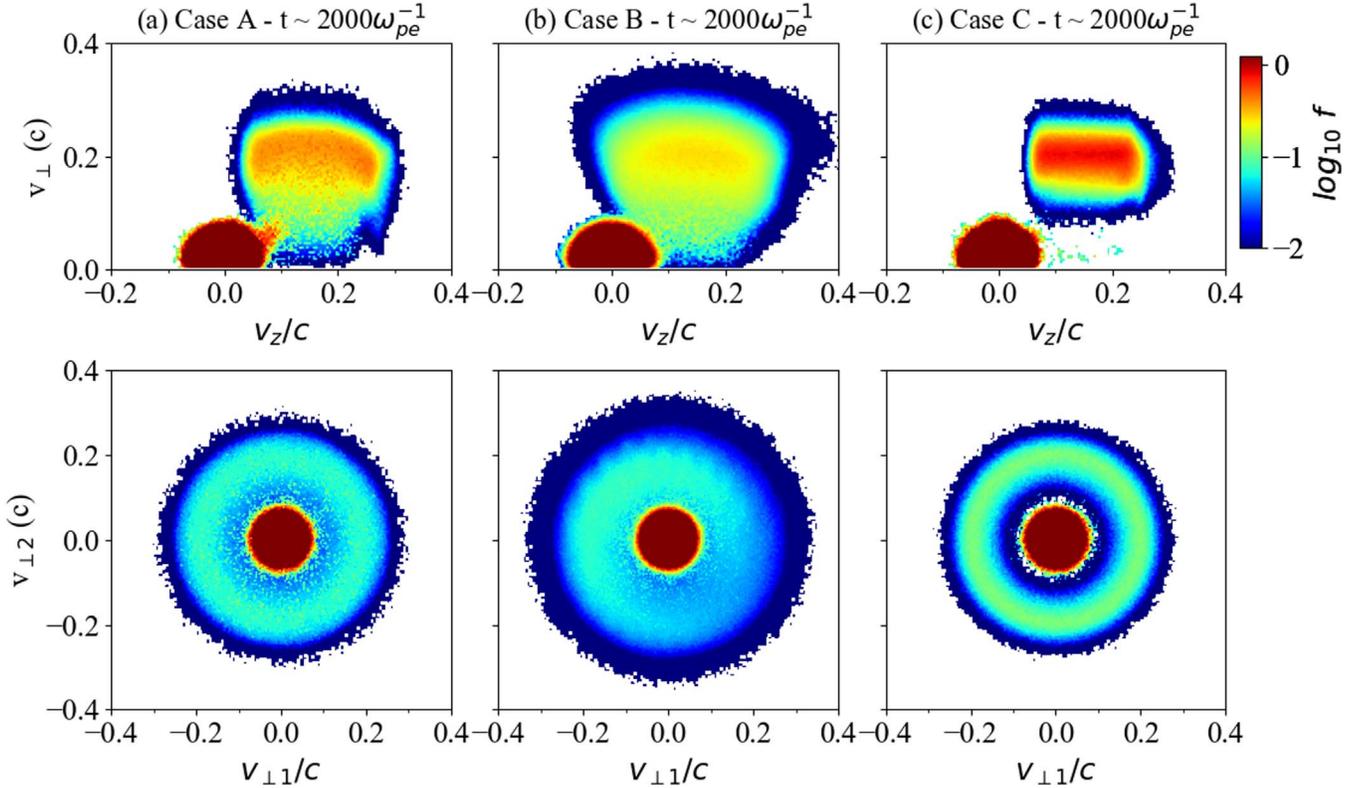

**Figure 2.** PIC-evolved EVDFs at the end of the simulation ($t = 2000\ \omega_{pe}^{-1}$) for cases A ($\omega_{pe}/\Omega_{ce} = 2.2$), B ($\omega_{pe}/\Omega_{ce} = 10$), and C ($\omega_{pe}/\Omega_{ce} = 1$). The upper panels illustrate the final stage of evolution in the $v_\perp - v_\parallel$ planes, while the lower panels show this in the $v_{1\perp} - v_{2\perp}$ planes. The video begins at time $t = 0\ \omega_{pe}^{-1}$, lasting until the simulation ends at $t = 2000\ \omega_{pe}^{-1}$. The real-time duration of the video is 10 s.
(An animation of this figure is available in the online article.)

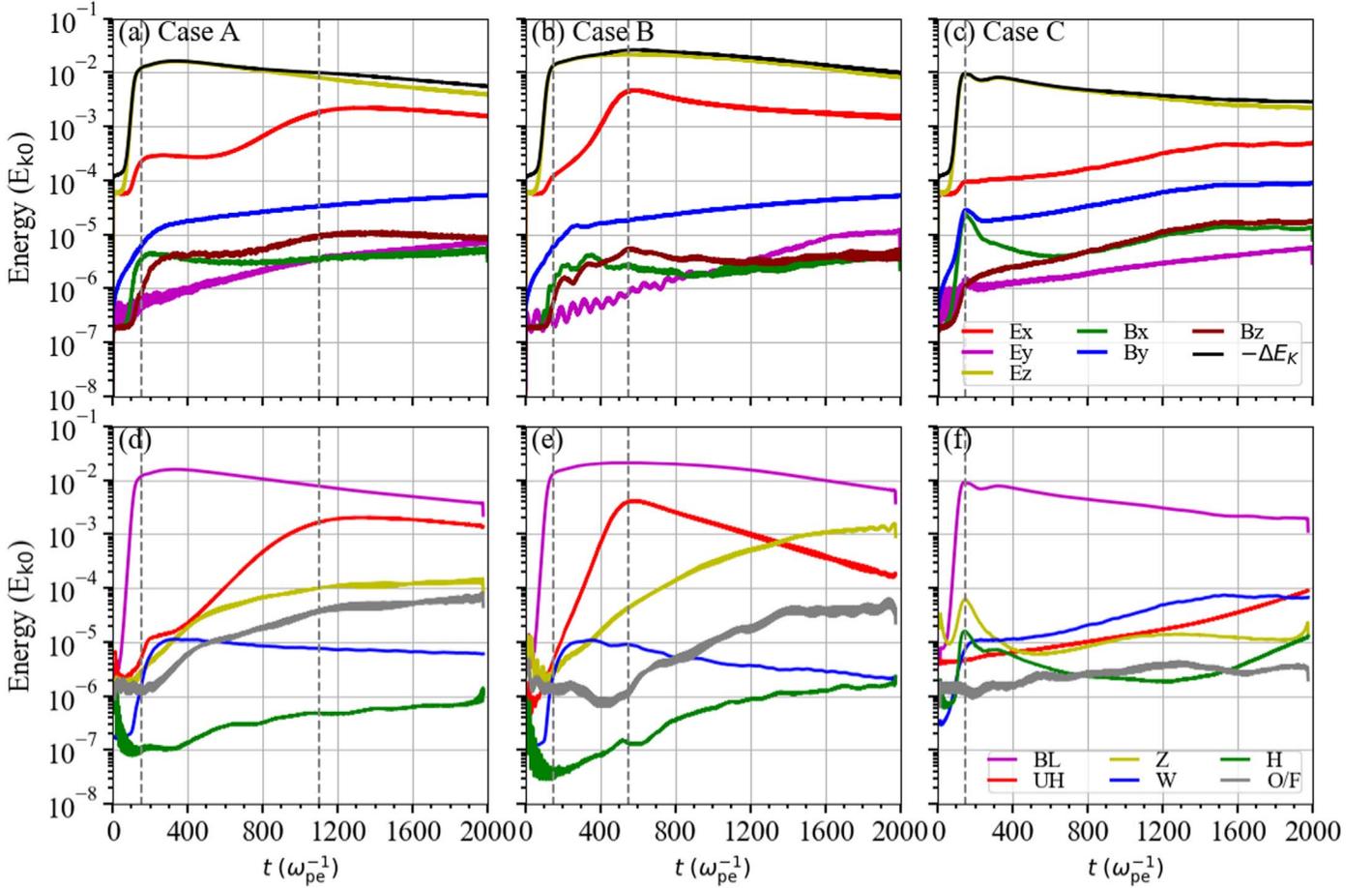

**Figure 3.** Upper panels: temporal evolution of energy changes for the six field components (*Ex*, *Ey*, *Ez*, *Bx*, *By*, and *Bz*) along with the negative variation of total electron energy ($-\Delta E_k$) for cases A ($\omega_{pe}/\Omega_{ce} = 2.2$), B ($\omega_{pe}/\Omega_{ce} = 10$), and C ($\omega_{pe}/\Omega_{ce} = 1$). Lower panels: temporal energy profiles of various wave modes (BL, UH, Z, W, O/F, and H), each normalized to the initial kinetic energy of energetic electrons ($E_{k_0}$). The vertical lines in different cases indicate the final times for stages I and II.

In Figure 3(a), we present the temporal evolution of the energy curves for the six electromagnetic field components alongside the negative variation of the total kinetic energy of all electrons ($-\Delta E_k$) for case A. Here, we divide the entire simulation into three main stages, based on the temporal energy profiles of the dominant field components for each mode. During the first stage ($\sim$0–150 $\omega_{pe}^{-1}$), the *Ez* component experiences significant growth, marking the early phase of wave excitation, with its maximum intensity accounting for a few percent of $-\Delta E_k$. In the second stage ($\sim$150–1100 $\omega_{pe}^{-1}$), *Ex* steadily grows to its peak energy, followed by a transition into the post-saturation stage, where the energy of other field components stabilizes.

The energy curves for each mode can be assessed using the Gaussian filter method (X. Zhou & S. Liu 2020; Z. Zhang et al. 2023). This method follows a systematic approach to analyzing the energy distribution of particular wave modes. Initially, the $\omega$ and k ranges of the targeted mode are identified based on the dispersion diagram plot (e.g., Figures 5 and 6). Next, a Gaussian profile is assumed for the wave distribution, where performing an inverse Fourier transform will result in the wave energy distribution in both time and space. The resulting energy is integrated spatially to generate each mode's temporal energy profile (e.g., see Figures 3(d)–(f)).

A detailed analysis of the energy of the dominating field component(s) of each mode reveals that *Ez* and *Ex* are primarily associated with the BL, UH, and Z modes, whereas the magnetic field components predominantly energize the W mode (see Figure 3(d)). The first stage in Figure 3(d) is primarily associated with the main BL mode. The second stage features the gradual rise and peak excitation of the UH, Z, and F modes. Notably, the W mode shows a nearly consistent increase from the start of the simulation up to about $\sim$250 $\omega_{pe}^{-1}$. After this point, it remains stable, with no significant changes, indicating that saturation has occurred. The H mode shows a slow and gradual rise from the beginning up to the end of the simulation, which appears to be insignificant.

Wave modes are identified using Fourier analysis on the obtained field data. Figure 4 and the accompanying movie illustrate the wave distribution map in $\mathbf{k}(k_\parallel, k_\perp)$-space, highlighting the strongest waves for specific $\mathbf{k}$. The movie further demonstrates the temporal evolution of this wave distribution. In Figure 5, we show the analysis of the $\omega$–$k$ dispersion across propagation angles ($\theta$, the angle between $\mathbf{k}$ and $\mathbf{B}_0$) at two different times of the simulation, revealing a significant increase in wave activity. The upper panel of Figure 5 illustrates the $\omega$–$k$ dispersion analysis for case A, including the analytical dispersion curves of the magnetoionic X, O, Z, and W modes, alongside the Langmuir or UH wave with thermal effects. By combining this figure with Figure 4, we can gain a complete understanding of the characteristics of each excited mode. The growth of $E_z$ and $E_x$ indicates the excitation of two primary electrostatic modes—the BL and UH modes, respectively, with BL being stronger and growing faster than UH. Previous studies have shown the excited modes under ring- or the beam-type EVDFs, where the bump-on-tail instability is driven by the beam (Y. Chen et al. 2022a) and the ECMI by the ring (Y. Chen et al. 2022b) components.

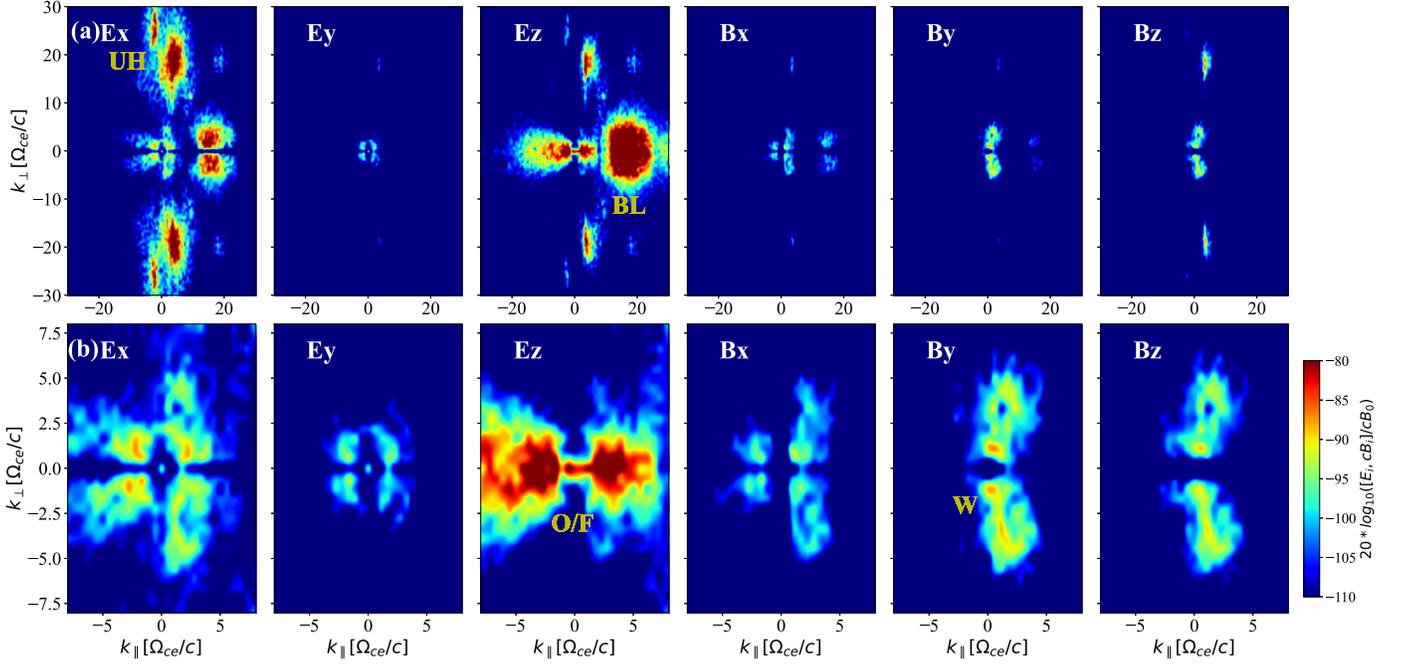

**Figure 4.** Upper panels: intensity maps of the six field components in the wavevector $k$-space for case A ($\omega_{pe}/\Omega_{ce} = 2.2$). Lower panels: a zoomed-in version of the intensified part of the top panel is presented. Some excited wave modes are shown in yellow within the figure. The video begins at time $t = 0\ \omega_{pe}^{-1}$, lasting until the simulation ends at $t = 2000\ \omega_{pe}^{-1}$. The real-time duration of the video is 10 s.
(An animation of this figure is available in the online article.)

In Figures 5(a)–(f), the upper panel shows the different excited modes observed in the earlier stage of the simulation (150–350 $\omega_{pe}^{-1}$), while the lower panel shows the modes in the later stage (1100–1300 $\omega_{pe}^{-1}$). The most prominent BL mode is evident in the $\omega - k$ range of [1.5, 2.5] $\Omega_{ce}$ and [10, 25] $\Omega_{ce}/c$, as represented in Figure 5(a). This mode shows small energy changes after 150 $\omega_{pe}^{-1}$ until the end of the simulation, as illustrated by the energy variation plots of the field components (Figure 3(d)). On the other hand, in Figure 5(b), there is no sign of the UH mode during the initial stage of the simulation. However, as time goes by, the UH mode begins to emerge within the range of [2.2, 2.8] $\Omega_{ce}$ and [−25, 25] $\Omega_{ce}/c$. The enhanced UH mode is locally distributed in $k$-space, as shown in Figure 4. The electrostatic UH mode should be mainly generated by the perpendicular evolution of the EVDFs, as demonstrated by the temporal profile growth of the UH mode from nearly 150 $\omega_{pe}^{-1}$ to 1100 $\omega_{pe}^{-1}$ of the simulation.

Among the presented modes near the plasma frequency $\omega_{pe}$ (i.e., the prominent BL and UH modes), there are other weaker modes worth noting. These include the small-$k$ Z mode and the large-$k$ backward-propagating Langmuir wave (L′), as depicted in Figure 5(a) at a later stage of the simulation. The Z mode (Figure 5(c)) is characterized as an electromagnetic mode with a phase velocity typically around or exceeding the light speed. This mode primarily manifests through the electric fields of $E_x$ and $E_z$. Its frequency ($\omega$) and wavevector ($k$) lie approximately within the intervals of [2–2.5] $\Omega_{ce}$ and [−10, 10] $\Omega_{ce}/c$, respectively. The excitation on the B dispersion map is mainly associated with the W mode. In Figure 5(d), the growth of the W mode is shown by $B_y$ within the ranges of [0, 0.9] $\Omega_{ce}$ and [−7, 7] $\Omega_{ce}/c$, lower than the characteristic frequency ratio of the simulation.

We also observed significant F and H plasma emissions, as respectively shown in Figures 5(f) and (e). The circular pattern indicates the excitation of H plasma emission within the frequency range of [3.5, 5] $\Omega_{ce}$ and the wavenumber range of [−4, 4] $\Omega_{ce}/c$. The perpendicular propagation of F emission spans the frequency range of [2.1, 2.3] $\Omega_{ce}$ and the wavenumber range of [−1, 1] $\Omega_{ce}/c$. At the end of our simulation, the F emission achieves an energy level of approximately $\sim 10^{-4} E_{k0}$ (see Figure 3(d)). The frequency of the F emission aligns well with the dispersion curves of the corresponding magnetoionic O mode (see Figure 5(f)), confirming that the observed F emission occurs in the O mode.

### 3.2. Wave Analysis for Cases with Different Frequency Ratios

This subsection provides additional PIC analysis for cases B ($\omega_{pe}/\Omega_{ce} = 10$) and C ($\omega_{pe}/\Omega_{ce} = 1$). Here, case B is divided into three stages: 0–150 $\omega_{pe}^{-1}$, 150–550 $\omega_{pe}^{-1}$, and 550–2000 $\omega_{pe}^{-1}$. In contrast, case C is divided only into two phases, due to less effective wave–particle interaction within the ranges of 0–150 $\omega_{pe}^{-1}$ and 150 $\omega_{pe}^{-1}$ up to the end of the simulation. The upper and lower panels of Figures 2(b)–(c), along with Figures 3(b)–(c) and 3(e)–(f), illustrate the PIC-evolved VDFs, the temporal energy profiles of various field components, and the energy of the dominant wave modes for cases B and C, respectively. The results of the $\omega - k$ dispersion analysis, highlighting various wave modes, are presented in Figures 6(a)–(f) across the upper and lower panels, respectively, for cases B and C.

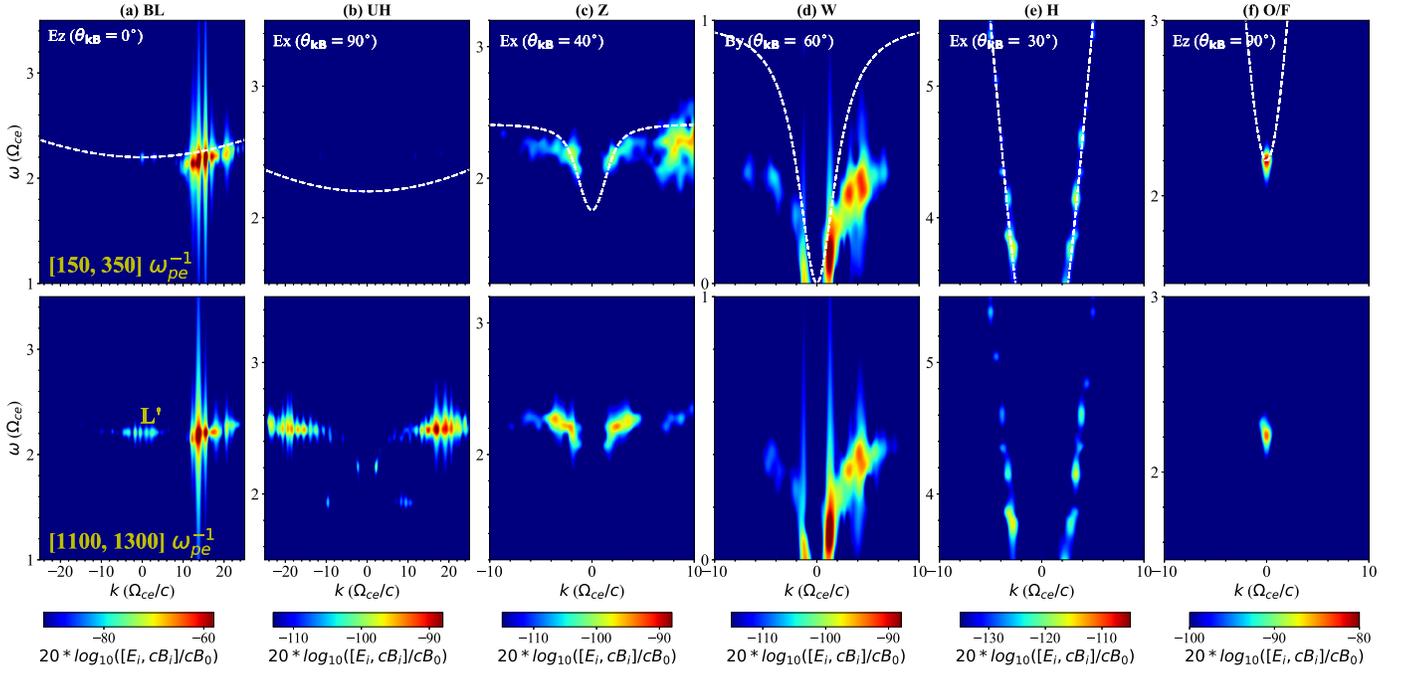

**Figure 5.** The dispersion diagrams of the field components for case A, which displays the BL (a), UH (b), Z (c), W (d), H (e), and O/F (f) modes. The upper panels show the time interval of the analysis at the initial stage of the simulation ($t = 150$–$350\omega_{pe}^{-1}$), while the lower panels show the same wave modes during a later time of the simulation ($t = 1100$–$1300\omega_{pe}^{-1}$). The upper part of each plot displays the specific field components and their corresponding propagation angles for the shown modes. The backward-propagating Langmuir wave (L') is produced and displayed during a later stage of the simulation. The overlaid lines represent dispersion curves derived from magnetoionic theory.

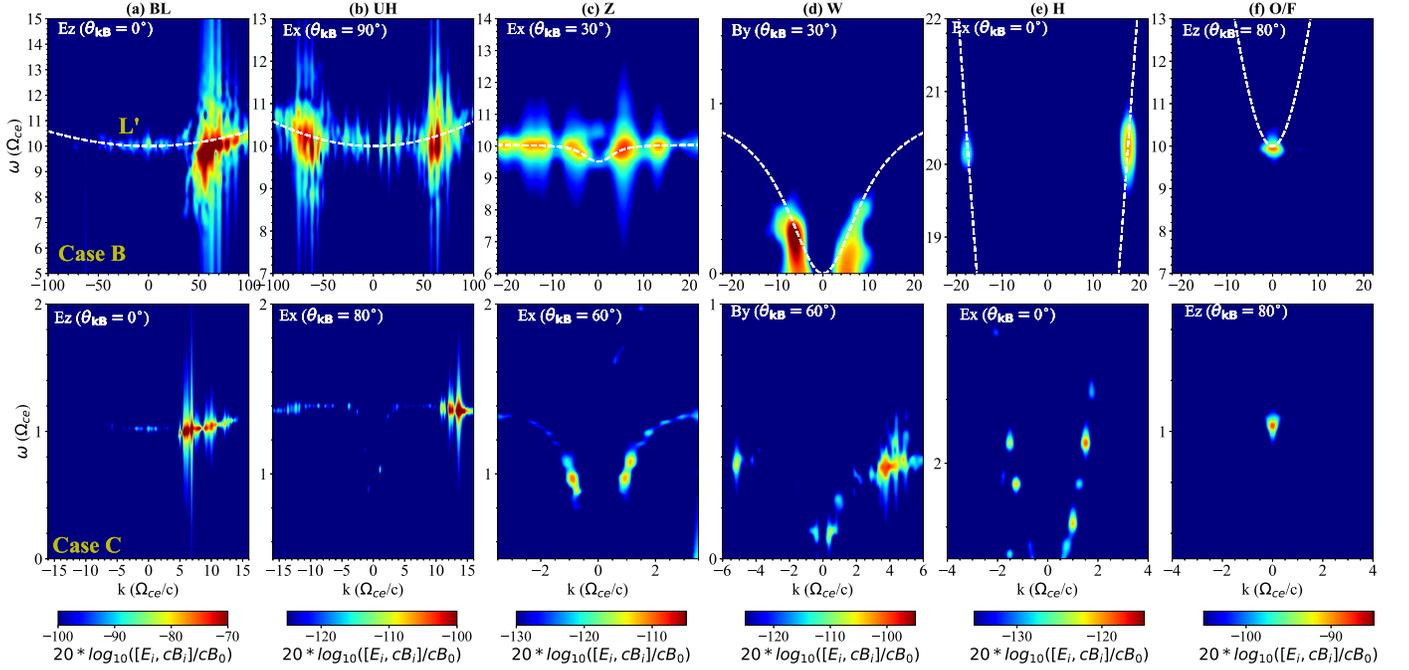

**Figure 6.** The same plot as Figure 5 but for specific simulation times for cases B (upper panels) and C (lower panels).

From the energy curves in Figures 3(d)–(f), it is clear that the durations of the linear stages in the BL mode are nearly identical across all three cases, with energy peaking around $\sim 0.01 E_{k_0}$ (at $t \sim 150\ \omega_{pe}^{-1}$). In contrast, the linear stages of the UH mode last about 1100 $\omega_{pe}^{-1}$ in case A, which is longer than case B ($\sim 550\ \omega_{pe}^{-1}$). Also, in case A, the energy peak goes up to about $\sim 0.002 E_{k_0}$, whereas in cases B and C, it reaches maximum values of $\sim 0.004 E_{k_0}$ and $\sim 10^{-4} E_{k_0}$, respectively. Also, the relevant wavenumbers of the BL and UH modes change from [−25, 25] $\Omega_{ce}/c$, [−100, 100] $\Omega_{ce}/c$, and [−16, 16] $\Omega_{ce}/c$ in cases A, B, and C, respectively (see Figures 5(a)–(b) and 6(a)–(b)). The energy excitation of the UH mode appears to increase in our findings. This increase is consistent across various cases, with case B—representing the highest frequency ratio—showing the highest UH energy excitation. In contrast, case C, which has the lowest frequency ratio, exhibits a weaker UH mode. Note that as the frequency ratio increases in different cases, it tends to speed up the excitation of the UH mode.

The W-mode energy growth increases as the frequency ratios decrease. In case C, the W mode reaches the highest energy, increasing up to approximately $\sim 7.5 \times 10^{-5} E_{k0}$, whereas in cases A and B, the energy almost reaches as high as about $\sim 10^{-5} E_{k0}$. The excited frequency of the W mode is less than unity for all three cases, where the reduced wavenumber range (from [−16, 16] $\Omega_{ce}/c$ down to [−6, 6] $\Omega_{ce}/c$) corresponds to decreasing frequency ratios in different cases.

The Z mode shows an upward trend in both cases A and B, with the highest slope observed in case B, peaking at a value of $\sim 1.7 \times 10^{-3} E_{k_0}$. In contrast, case A only reaches up to an order of $\sim 2 \times 10^{-4} E_{k_0}$. In the first two cases, this mode shows an upward trend, unlike case C, which peaks early in the simulation at $\sim 6 \times 10^{-5} E_{k_0}$, around $t \sim 150\, \omega_{pe}^{-1}$. The spectral intensity of the Z mode weakens as the frequency ratio decreases, resulting in a narrower frequency range.

The intensity of the F plasma emissions is relatively high for cases A and B, where the average intensity measures roughly around $\sim 10^{-4} E_{k_0}$ and $\sim 9 \times 10^{-5} E_{k_0}$, respectively. We also have a weaker F mode, with the peak energy reaching up to $\sim 5 \times 10^{-6} E_{k_0}$ for case C. The O/F mode is primarily driven by the parallel electric field $E_z$ in all cases. The peak energy of the H emission changes in different scenarios. In cases A and B, the peak energy level increases to approximately $\sim 2 \times 10^{-6} E_{k0}$. This further increases to about $\sim 1.5 \times 10^{-5} E_{k_0}$ in case C, where the frequency ratio is equal to unity. Note that for the first two cases (A and B), the strength of the F mode in the crescent-shaped EVDF is greater compared to the H mode, while for case C, the H mode is stronger than the F mode at the end of the simulation.

In the earlier-studied ring–beam case, the BL mode primarily propagated along the parallel direction within a limited range of $k_\parallel$ and $k_\perp$. In contrast, the UH mode grew along blob regions in $\mathbf{k}$-space, clustered by an oblique line corresponding to harmonic numbers, consistent with the linear wave-growth-rate analysis calculated by Z. Zhang et al. (2023). Here, the results of cases A and B can be compared with the ring–beam case (Z. Zhang et al. 2023), but the intensity of the excited F and H modes in case C is closer to the pure-ring case, where H is the dominant mode (Y. Chen et al. 2022b). In case B, the intensity of the fundamental mode is greater than what has been previously reported for the pure-ring (Y. Chen et al. 2022b) and pure-beam (Y. Chen et al. 2022a) cases, and even approximately twice as high as the ring–beam cases (Z. Zhang et al. 2023), considering similar frequency ratios.

## 4. Summary and Discussion

In this study, we have employed the fully kinetic electromagnetic PIC code to simulate the emission properties of crescent-shaped EVDFs under different frequency ratios. We have performed a comparative analysis of wave emissions and intensities across three scenarios: the frequency ratio was defined as 2.2, 10, and 1 (labeled as cases A, B, and C, respectively), to explore various aspects of these crescent-shaped EVDFs.

In the first case of our study (case A), we employed an EVDF and a frequency ratio ($\omega_{pe}/\Omega_{ce} = 2.2$) similar to that discussed in X. Yao (2022b). However, the key differences in the initial simulation setup and corresponding results highlight the distinct nature of our findings. While X. Yao (2022b) used a mass ratio of 100 in their simulations, we employed the more realistic physical value of 1836. Furthermore, we significantly increased the number of particles per cell by a factor of 10, leading to higher resolution and improved accuracy in our simulations. We also set the density ratio of the energetic to background electrons at 0.01, in contrast to the much higher 0.5 used in X. Yao (2022b). This adjustment reflects a more generalized and realistic scenario, where the population of energetic electrons is relatively small compared to the background ones.

These differences resulted in a high impact on the wave emissions observed in our simulations, despite both studies employing crescent-shaped EVDFs under the same frequency ratio. Whereas X. Yao (2022b) focused on ECMIs leading to the generation of multiple-harmonic electron cyclotron waves, our results demonstrated the significant excitation of BL and UH modes. The prominently excited BL and UH modes result in the observed F and H plasma emissions in our simulations, where their energy levels can rise to relatively high values.

In the second case of the presented study (case B), our excited modes for the crescent shape are similar to the previously published ring–beam case with the same frequency ratio ($\omega_{pe}/\Omega_{ce} = 10$; Z. Zhang et al. 2023). In both studies, efficient excitation of the BL and UH modes was observed, along with notable plasma emissions. Here, for a better and more comprehensive understanding, we have also examined the crescent-shaped VDF from multiple perspectives, by rotating the crescent components of the VDFs (a rotated version of Figure 1(b) is not shown). Essentially, the crescent-shaped VDFs can be roughly considered as parts of a ring with more intense portions than the other areas. Based on our analysis, the ring components of ring–beam VDFs serve as locations that excite certain modes, while in the crescent-shaped VDF, the intensified part only alters the intensity of the excited modes, without causing the excitation of new wave modes. Our findings suggest that the similar shape of the VDF in crescents and ring beams primarily influences the free energy distribution of the energetic electrons but not the fundamental mechanisms driving wave excitation and plasma emission.

In case C of our simulations, where the frequency ratio $\omega_{pe}/\Omega_{ce} = 1$, we observed significant excitation of the F and H emissions by the crescent-shaped EVDFs, with the H mode being particularly dominant. The frequency ratio around or lower than case C is often observed at heights within one solar radius or less in active regions where large or highly twisted magnetic fields are common (S. Regnier 2015; D. E. Morosan et al. 2016; M. Yousefzadeh et al. 2021). Our research suggests that when a crescent-shaped EVDF is present in an environment with $\omega_{pe}/\Omega_{ce} = 1$, there is a high chance for exciting H emission.

The energy conversion rates from energetic electrons into different modes vary significantly across cases. The BL mode exhibits a conversion rate of approximately 1%–2%, while the UH mode shows a decrease from around 0.2% in case A to 0.4% in case B, and further drops to about 0.01% in case C. In comparison, the maximum conversion rates for the W and Z modes remain at less than 0.01% and approximately 0.2%, respectively, across the lowest and highest frequency ratios. The fundamental (O/F) mode can achieve conversion rates up to nearly 0.01% in both cases A and B, whereas the H mode reaches a maximum of 0.0015% only in case C.

These results highlight the complex energy-exchange mechanisms driven by the interaction of crescent-shaped EVDFs with plasma instabilities. The free energy of these distributions is efficiently converted into specific wave modes, with the BL and UH modes dominating the electric fields and the W mode channeling energy through the magnetic fields, influenced by the specified frequency ratio. Additionally, the residual energy is redistributed to the thermal energy of the ambient plasma through wave dissipation and particle scattering, reflecting the complex dynamics of wave–particle interactions under the given plasma conditions.

These instabilities possibly result in the efficient excitation of the BL mode, leading to the generation of F and H emissions. Based on the theoretical framework of plasma emission, the F emission is achieved as the BL mode scatters or decays into ion acoustic waves. The H emissions are generated when the BL mode coalescences with L′ or/and the coalescences of forward- and backward-propagating UH modes, as proposed by Y. Chen et al. (2022a) and Y. Chen et al. (2022b). These findings suggest that the crescent-shaped EVDF can efficiently excite F and H plasma emissions under specific frequency ratio conditions.

## Acknowledgments

The research results are sponsored by the China/Shandong University International Postdoctoral Exchange Program and NSFC grants (12203031, 42074203, and 42127804). The author is grateful for the open-source Vector Particle In Cell (VPIC) code provided by Los Alamos National Laboratory (LANL) and the Super-Cloud Computing Center (BSCC, http://www.blsc.cn/) for providing computational resources. The author gratefully acknowledges Prof. Yao Chen (SDU) for his guidance and helpful discussions.

## ORCID iDs

Mehdi Yousefzadeh  https://orcid.org/0000-0003-2682-9784

## References

Bessho, N., Chen, L.-J., & Hesse, M. 2016, GeoRL, 43, 1828
Bowers, K. J., Albright, B. J., Bergen, B., et al. 2008a, in SC'08: Proc. of the 2008 ACM/IEEE Conf. on Supercomputing, 63 (IEEE: Piscataway, NJ), 1
Bowers, K. J., Albright, B. J., Yin, L., et al. 2008b, PhPl, 15, 055703
Bowers, K. J., Albright, B. J., Yin, L., et al. 2009, JPhCS, 180, 012055
Burch, J., Dokgo, K., Hwang, K., et al. 2019, GeoRL, 46, 4089
Chen, Y., Zhang, Z., Ni, S., et al. 2022a, ApJL, 924, L34
Chen, Y., Zhang, Z., Ni, S., et al. 2022b, PhPl, 29, 112113
Dokgo, K., Hwang, K., Burch, J. L., et al. 2019, GeoRL, 46, 7873
Dory, R. A., Guest, G. E., & Harris, E. G. 1965, PhRvL, 14, 131
Ginzburg, V. L., & Zhelezniakov, V. V. 1958, SvA, 2, 653
Graham, D. B., Vaivads, A., Khotyaintsev, Y. V., et al. 2018, JGRA, 123, 2630
Henri, P., Sgattoni, A., Briand, C., Amiranoff, F., & Riconda, C. 2019, JGRA, 124, 1475
Lapenta, G., Berchem, J., Zhou, M., et al. 2017, JGRA, 122, 2024
Lu, Q. M., Wu, C. S., & Wang, S. 2006, ApJ, 638, 1169
Morosan, D. E., Zucca, P., Bloomfield, D. S., et al. 2016, A&A, 589, L8
Ni, S., Chen, Y., Li, C., et al. 2020, ApJL, 891, L25
Regnier, S. 2015, A&A, 581, A9
Thurgood, J. O., & Tsiklauri, D. 2015, A&A, 584, A83
Umeda, T. 2010, JGRA, 115, A01204
Wild, J. P. 1950, AuSRA, 3, 541
Wild, J. P., & McCready, L. L. 1950, AuSRA, 3, 387
Wild, J. P., Murray, J. D., & Rowe, W. C. 1954, AuJPh, 7, 439
Wu, C. S., Wang, C. B., Wu, D. J., & Lee, K. H. 2012, PhPl, 19, 082902
Yao, X., Muñoz, P. A., & Büchner, J. 2022a, PhPl, 29, 022104
Yao, X., Muñoz, P. A., Büchner, J., et al. 2022b, ApJ, 933, 219
Yousefzadeh, M., Ning, H., & Chen, Y. 2021, ApJ, 909, 3
Zhang, Z., Chen, Y., Ni, S., et al. 2022, ApJ, 939, 63
Zhang, Z., Chen, Y., Ni, S., et al. 2023, PhPl, 30, 122106
Zhou, X., Muñoz, P. A., Büchner, J., & Liu, S. 2020, ApJL, 891, 92